\documentclass[letterpaper]{tMOP2e}

\usepackage{amsmath}
\usepackage{verbatim}
\usepackage{xcolor}

\begin{document}
\doi{10.1080/0950034YYxxxxxxxx}
 \issn{1362-3044}
\issnp{0950-0340} \jvol{00} \jnum{00} \jyear{2013} \jmonth{10 January}

\markboth{Mohammad Mirhosseini, Brandon Rodenburg, Mehul Malik, and Robert W. Boyd}{Journal of Modern Optics}

\title{\itshape Free-space communication through turbulence: a comparison of plane-wave and orbital-angular-momentum encodings}

\author{Mohammad Mirhosseini$^{a}$$^{\ast}$\thanks{$^\ast$Corresponding author. Email: mirhosse@optics.rochester.edu
\vspace{6pt}}, Brandon Rodenburg$^{a}$, Mehul Malik$^{a}$ and Robert W. Boyd$^{a,b}$\\\vspace{6pt}  
$^{a}${\em{The Institute of Optics, University of Rochester. 320 Wilmot BLDG, 275 Hutchison Rd, Rochester NY 14627, USA}};
$^{b}${\em{Department of Physics, University of Ottawa, Ottawa, ON K1N 6N5, Canada}}
\\\vspace{6pt}\received{v3 released Februrary 2013} }

\maketitle

\begin{abstract}

Free-space communication allows one to use spatial mode encoding, which is susceptible to the effects of diffraction and turbulence. Here, we discuss the optimum communication modes of a system while taking such effects into account. We construct a free-space communication system that encodes information onto the plane-wave (PW) modes of light. We study the performance of this system in the presence of atmospheric turbulence, and compare it with previous results for a system employing orbital-angular-momentum (OAM) encoding. We are able to show that the PW basis is the preferred basis set for communication through atmospheric turbulence for a large Fresnel number system. This study has important implications for high-dimensional quantum key distribution systems.
\end{abstract}

\bigskip

\section{Introduction}
\noindent Free-space optical links provide an easy means of high-bit-rate communications for line-of-sight systems. Furthermore, these links require a much simpler infrastructure as compared to fiber-based systems.  The majority of the proposed applications of free-space optical communications use polarization, wavelength, or the time degree of freedom of the optical fields for encoding information \cite{Chan:2006vr,Kerr:1970vo}. Recently, there have been a number of studies suggesting the use of spatial profile of optical fields as an extra degree of freedom for encoding information \cite{Boyd:2011wv,Miller:2000tw,malik2012influence}. It has been demonstrated that the use of spatial modes as an extra layer of multiplexing can substantially increase the bit rate of a classical communication link \cite{Wang2012}.  In addition, the large Hilbert space of spatial modes makes quantum key distribution (QKD) systems more tolerant to eavesdropping errors \cite{Bourennane:2001cp}. 

Orbital angular momentum (OAM) is often suggested as a preferable set of spatial modes for free-space communication. It has been argued that, in principle, there is no limit on the number of bits of information that can be carried by a single photon using this encoding scheme \cite{Boyd:2011wv, Wang2012}. In practice, however, the number of spatial modes that can be communicated in any free-space system is always limited by the undesired effects of diffraction and atmospheric turbulence. In the presence of these adverse effects, the choice of optimal encoding basis is not obvious. 
In this paper, we briefly review the effects of diffraction on the propagation of scalar fields. This analysis suggests that the OAM modes form an optimal basis for optical systems with rotational symmetry. However, in practical systems with large apertures, diffraction does not play a major role, and the channel capacity is mostly limited by turbulence. Motivated by this argument, we experimentally investigate the propagation of an alternative group of spatial modes in the presence of atmospheric turbulence.  This basis comprises an orthonormal set of uniform beams with tilted wavefronts, also known as plane wave (PW) modes. The set of PW modes forms a large orthonormal basis set that can be used for encoding information. Furthermore, the PW modes can easily be separated by using a single lens at the receiving aperture. Mode decomposition analysis is used to perform a quantitative comparison of the performance of PW and OAM modes under the adverse effects of turbulence. The results suggest that the PW basis is more robust against atmospheric turbulence, closely confirming the theoretical predictions of Boyd \emph{et al.} \cite{boyd2011influence}.
\section{Diffraction and communication modes}

\begin{figure}[h]
	\centerline{\includegraphics[scale=0.65]{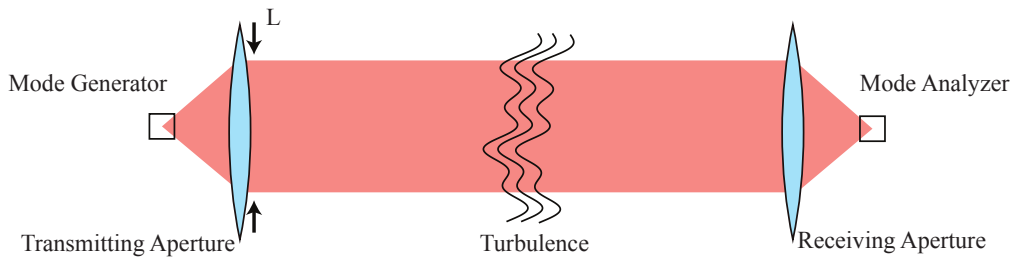}}
	\caption{Schematic diagram of a free-space communication link.}
	\label{fig:link}
\end{figure}

Consider a prototypical free-space communication link as depicted in Fig. \ref{fig:link}. Due to diffraction, a spatially confined beam generated in the transmitting aperture spreads upon propagation. The area of the beam over which most of its energy is confined can be calculated at any plane using diffraction theory. However, after propagating beyond the near field, the beam has long tails which continue toward infinity in the transverse plane\cite{Goodman:Fourier}. Since the receiving aperture has a finite size, a portion of energy of the beam will be lost in the detection process. Additionally, the strength of this induced loss is dependent on the form of the transmitted field. As a result, a set of initially orthogonal fields in the transmitting aperture will no more form an orthogonal set at the receiving aperture, and it is not passible to perfectly discriminate between them. The problem of finding the optimum set of modes for transmitting energy and information from one finite aperture to another has long been investigated in the context of apodization theory \cite{Slepian:1965wz} and the theory of communication modes \cite{Miller:2000tw}.

The input-output characteristics of a communication link can be described using the Rayleigh-Sommerfeld diffraction formula, which describes the free-space propagation of an arbitrary scalar field
\begin{equation}\label{eq:RS}
 \Psi_{out}(\mathbf{r}_R)=\int \Psi_{in}(\mathbf{r}_T) K(\mathbf{r}_T,\mathbf{r}_R)d^2\mathbf{r}_T,
\end{equation}
in which \(\Psi_{in}(\mathbf{r}_T)\) and  \(\Psi_{out}(\mathbf{r}_R)\) are the electric fields in the transmitting and the receiving apertures. The propagation kernel \(K(\mathbf{r}_T,\mathbf{r}_R)\) can be written as \cite{Goodman:Fourier}
 \begin{equation}
K(\mathbf{r}_T,\mathbf{r}_R)=-\frac{1}{2\pi}\frac{\partial}{\partial z}\frac{\exp(ik\mid \mathbf{r}_T-\mathbf{r}_R \mid)}{\mid \mathbf{r}_T-\mathbf{r}_R \mid}.
\end{equation}

Eq. (\ref{eq:RS}) indicates that the fields in the two apertures are related via a linear transformation. Using Dirac notation, this transformation can be written simply as \(\mid \Psi_{out}\rangle=\hat{P}\mid \Psi_{in}\rangle\), where \(\hat{P}\) is the propagation operator.
 Note that the propagation kernel as defined above represents a unitary transformation. In order to calculate the detected field, we must take into account the finite size of the apertures. This imposes the condition \(K(\mathbf{r}_T,\mathbf{r}_R)=0\) for  \(r_R>R_1\) and \(r_T>R_0\). As a result of adding this constraint, the propagator is no longer a unitary operator. A non-unitary transformation does not preserve inner products. Hence a set of orthogonal modes in the transmitting aperture are not necessarily orthogonal in the receiving aperture upon propagation
 \begin{equation}
\langle \Psi_{m-out}\mid \Psi_{n-out}\rangle = \langle \hat{P}\Psi_{m-in}\mid\hat{P}\Psi_{n-in}\rangle = \langle \Psi_{m-in}\mid \hat{P}^\dagger \hat{P}\mid\Psi_{n-in}\rangle \neq 0.
\end{equation}
Note that since \(\hat{P}\) is not unitary, \(\hat{P}^\dagger \hat{P}\) need not constitute the identity matrix. However, it is still possible to find a specific set of orthogonal modes in the transmitting aperture which remain orthogonal upon propagation. These modes can be found by performing a singular value decomposition (SVD) of the propagator operator. This can be formally expressed as
 \begin{equation}
 \hat{P}^\dagger \hat{P}\mid\Phi_{n}\rangle=\mid\lambda_n\mid^2\mid\Phi_{n}\rangle.
 \label{eq:SVD}
 \end{equation}
The functions \(|\Phi_{n}\rangle\) are sometimes called the communication modes of the system \cite{Miller:2000tw,Martinsson2007}. The coefficient \(\mid\lambda^2_n\mid\) is a coupling coefficient and is equal to the portion of the energy of the corresponding mode which falls within the receiving aperture\cite{Miller:2000tw,Shapiro:2009vc}. 

The SVD procedure guarantees that the communication modes form a complete orthonormal set in the receiving aperture. 
\begin{equation}
\langle \Psi_{m-out}\mid \Psi_{n-out}\rangle = \langle \Psi_{m-in}\mid \hat{P}^\dagger \hat{P}\mid\Psi_{n-in}\rangle = \mid\lambda_n\mid^2 \langle \Psi_{m-in}\mid\Psi_{n-in}\rangle = \mid\lambda_n\mid^2 \delta_{mn}.
\end{equation}
Therefore, these modes form a preferred set for free-space communication. In order to find the exact form of the communication modes, the SVD procedure has to be performed on the propagator using the exact shapes and sizes of the apertures, as well as the distance between them. Considering the fact that the transformation kernel includes a rapidly oscillating term, the SVD procedure is a numerically demanding task for any system with a dimension larger than a few optical wavelengths. In the rest of this section, we utilize a combination of symmetry considerations and approximations to find an approximate solution for the communication modes of a realistic system. 

A real-world communication link usually consist of circular components such as lenses, apertures, and mirrors. In this situation, the rotational symmetry of the system can be exploited to find analytical solutions for Eq. (\ref{eq:SVD}). Assuming paraxial Fresnel diffraction, it can be shown that the solutions have the form 
 \begin{equation}
  \label{eq:PS}
 \Phi_{n, \ell}(\rho,\phi)=\frac{R_{n, \ell}(\rho)}{\sqrt{\rho}}e^{i\ell\phi},
\end{equation}
where the radial parts \(R_{n, l}(\rho)\) are known as the generalized prolate-spheroidal functions \cite{Slepian:1964ww, Slepian:1965wz}. The azimuthal dependence \(e^{il\phi}\) suggests that these solutions are eigenfunctions of the orbital angular momentum operator \cite{Allen1992}. The eigenfuntions of Eq. (\ref{eq:PS}) provide a complete two dimensional basis set for free-space communication between two circular apertures. In practice, however, it is easier and more common to create optical fields which still retain the vortex feature of \(e^{il\phi}\) , while using the radial profile of a Laguerre-Gaussian or a top-hat beam. This is because the functional dependence of the solutions on the radial and the azimuthal coordinates is separable, and any two functions with different OAM indices maintain their orthogonality upon propagation, regardless of their radial form. 

The logic presented above suggests that OAM modes are the natural choice for encoding information in a free-space optical link. However, it can be shown that for an optical system with sufficiently large apertures, finding communication modes and their coupling coefficients is a nearly degenerate problem \cite{Shapiro1974}. This means that for a given coupling coefficient \(\lambda_n\), there exist multiple functions \(\Psi_n\) satisfying Eq. (\ref{eq:SVD}). The number of these modes is approximately equal to the Fresnel number product of the system \(N_F = \frac{A_R A_T}{(\lambda L)^2}\), and their coupling coefficients are almost equal to unity \cite{Shapiro1974, Shapiro:2009vc}. This specific subset of the eigenfunctions is sometimes known as the degenerate communication modes. It is easy to see that, as a consequence of degeneracy, any linear superposition of the degenerate communication modes is itself an eigenfunction of Eq. (\ref{eq:SVD}). This allows us to use a linear transformation to map the degenerate communication modes to a new set of orthonormal modes. One such basis that can be formed using a superposition of the communication modes of Eq. (\ref{eq:PS}) is the basis of PW modes.

In the present article, we explore the possibility of employing an encoding scheme based on the use of the PW modes of light. This study is partly motivated by the fact that the generation and separation of PW modes are simple as compared to those of OAM modes. OAM modes are conventionally generated using high-resolution spatial light modulators \cite{rodenburg2012influence} or a series of forked holograms \cite{Wang2012}, whereas to generate a PW mode, one has to simply add a wavefront tilt to a top-hat beam. Similarly, sorting OAM modes needs carefully crafted custom optical elements \cite{Lavery2012, OSullivan:2012gj}, which are very sensitive to misalignment \cite{Lavery2011}. The process of separating PW modes, on the other hand, can be achieved with a single lens. This lens transforms different PW modes to spatially separated spots in its focal plane.

\section{Effects of turbulence on PW and OAM modes}

 In a classical communication system, aberrations induced by turbulence on the transmitted modes result in a spread of the detected modes in the receiver, increasing the cross-talk and thus reducing the channel capacity of the system \cite{rodenburg2012influence,Roux:2011dr}. In a QKD system, the turbulence induced loss of quantum coherence results in an increase of the error rate which can compromise the integrity of the protocol \cite{malik2012influence}. Here we consider the role of atmospheric turbulence on a free-space communication system that employs PW encoding at high light levels. The connection between quantum and classical properties can be understood by considering the fact that photons are units of excitation of the modes of light. Therefore the mode mixing introduced by the turbulence affects both classical and QKD systems in identical ways.  
 
 In our system, the transmitting aperture is imaged onto the receiving aperture. This situation represents the \(N_F \to \infty\) limit in which the effects of diffraction can be safely ignored. For such a system, turbulence can be modeled as a thin phase-screen in the transmitting aperture \cite{Young:1974vp}. This single phase-screen approximation is valid as far as the aberrations introduced by the turbulence are not too large. The phase aberrations can be described by normal random variables characterized by the quantity \(\langle [{\phi(r_1)-\phi(r_2)}]^2\rangle\). This quantity is known as the phase structure function and can be evaluated using Kolmogorov turbulence theory to give the result
\begin{equation}
\langle [{\phi(r_1)-\phi(r_2)}]^2\rangle=6.88\left |{\frac{x_1-x_2}{r_0}}\right |^{5/3}.
\end{equation}
The parameter \(r_0\) is known as Fried's coherence parameter and is a measure of the length of correlation of the phase aberrations \cite{FRIED:1965gu}. 
\begin{figure}[h]
	\centerline{\includegraphics[scale=1.0]{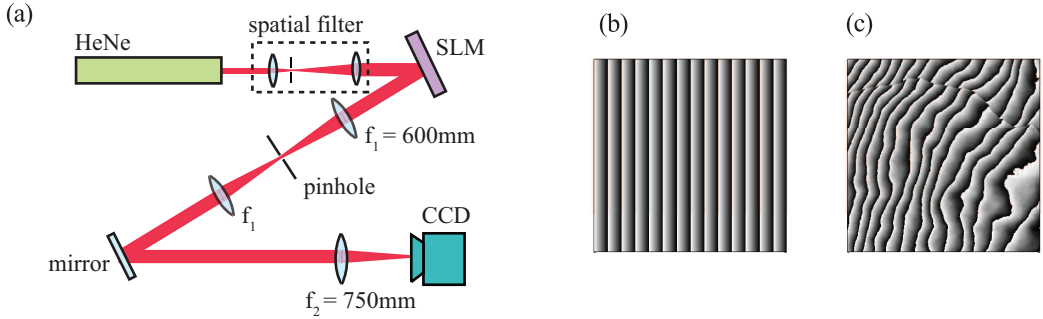}}
	\caption{(a) A PW mode is prepared by using a phase grating, of the sort shown in (b). This procedure is realized through use of an SLM illuminated by an expanded He-Ne laser beam. The first order diffracted beam is imaged by a 4-f system onto the receiving aperture, where a 750 mm lens separates the modes. (c) Thin-phase turbulence is added to the phase grating. The first-order diffracted beam in this case represents a PW mode after propagation through turbulent atmosphere}
	\label{fig:setup}
\end{figure}

Using the model of phase aberrations introduced above, Boyd \emph{et al.} have analyzed the effects of turbulence on the propagation of PW modes \cite{boyd2011influence}. The PW modes considered here are confined to a finite square aperture and have tilts only in one dimension. These modes as launched by the transmitter can be represented as
\begin{equation}
A(x,y)=A_0W(x/L)W(y/L)e^{im\frac{2\pi x}{L}}
\end{equation}
where \(A_0\) is the field amplitude, \(W(\xi)\) is the aperture function defined so that \(W(\xi) = 1\) for \(\mid\xi\mid \leq 1/2\) and zero otherwise, and \(m\) is the mode index of the launched field. The theoretical analysis in Ref. \cite{boyd2011influence} suggests that the turbulence-induced cross-talk in the detection of these modes can be calculated by evaluating the integral
\begin{equation}\label{PWnumerical}
\langle s_{\Delta} \rangle= 8\int_0^{1/2} d\eta {\left( {\frac{1}{2} - \eta}\right)} e^{-3.44(\eta L/r_0)^{5/3}} \cos(4\pi\Delta\eta)
\end{equation}
where \(L\) is the width of the rectangular transmission aperture, and \(\langle s_{\Delta}\rangle \) is the conditional probability of detecting a photon in the PW mode \(m+\Delta\), given that the photon was sent in the PW mode \(m\). The integration is performed over the normalized transverse coordinate \(\eta\) . Since the integral in Eq. (\ref{PWnumerical}) is only an explicit function of \(L/r_0\), the cross-talk probability can be quantified using this single parameter. 

A similar approach has previously been used to analyze the effects of turbulence on propagation of OAM states \cite{Paterson:2005km,Tyler2009,Gbur2008}. A numerical comparison of the performance of PW modes versus OAM modes has been done in Ref. \cite{boyd2011influence}. The authors have concluded that the PW encoding is less quickly degraded by a factor of about three. In this paper, we experimentally verify this result.

 Fig. (\ref{fig:setup}) shows the schematic diagram of our experiment. Spatially collimated light from a He-Ne laser illuminates a spatial light modulator (SLM), which is utilized along with a 4f system and an aperture to generate different PW modes. Since the turbulence is modeled by a single phase-screen in the transmitting aperture, the same SLM is used to impress Kolmogorov phase aberrations onto the beam . A 750 mm lens is employed at the receiving aperture to sort the different PW modes into spatially separated spots in its focal plane. A CCD is used for recording the intensity profile of the sorted modes in this plane. 
 \begin{figure}[h]
	\centerline{\includegraphics[scale=0.55]{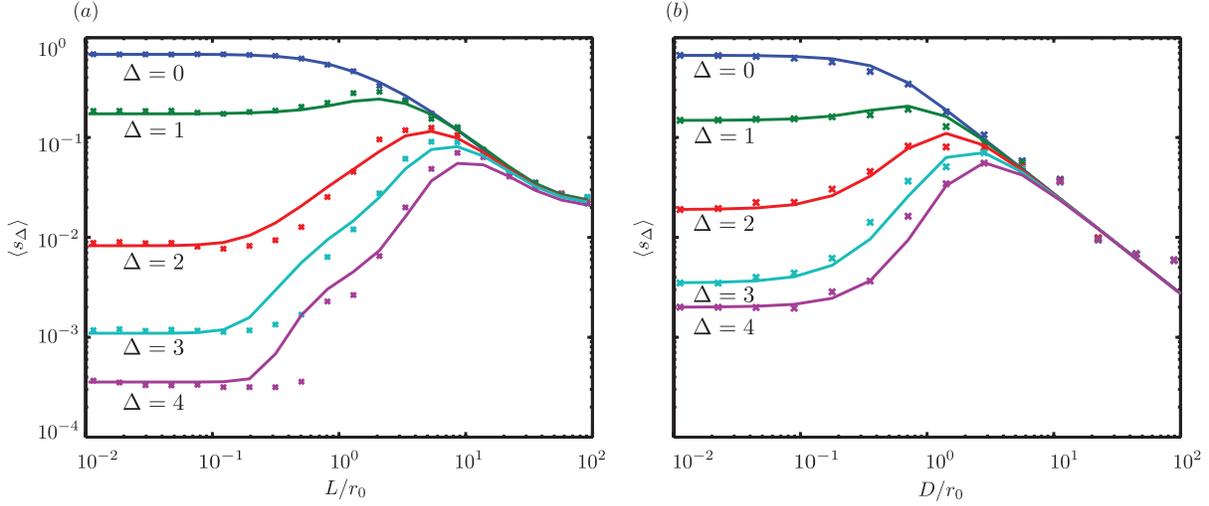}}
	\caption{(a) Average cross-talk values \(\langle s_\Delta\rangle\) for PW modes are plotted as a function of turbulence level \(L/r_0\) for the \(m=0\) input. The experimental data is presented by crosses whereas the theoretical predictions are shown by the solid lines. The plots  level off to constant values at the limit of very little turbulence due to the residual cross-talk arising from our sorting technique. (b) The theoretical and experimental values of \(\langle s_\Delta\rangle\) for an \(\ell=0\) OAM mode as reported in Ref. \cite{rodenburg2012influence}.}
	\label{fig:comparison}
\end{figure}

When the turbulence phase-screen is not present, each mode will map to a diffraction-limited spot on the CCD's screen. We divide the area on the CCD into non-overlapping adjacent spatial bins which correspond to the central positions of these spots. In the presence of turbulence, each mode will form a random shape on the screen. The amount of beam power falling within a spatial bins is proportional to the value of \(\langle s_{\Delta} \rangle\) for a given input mode. We have measured these values in 51 different regions corresponding to \(\mid \Delta \mid \leq 25\). A range of turbulence levels characterized by \(L/r_0 \in [10^{-2} :10^2] \) was tested. For each value of \(L/r_0\), the results are averaged over 100 phase-screens. To stay within the spatial bandwidth of the SLM, the minimum value of \(r_0\) was chosen to be sufficiently large . 

The theory presented in Ref. \cite{boyd2011influence} dictates that the average cross-talk values \(\langle s_{\Delta} \rangle\) are independent of the transmitted mode index \(m\). We have verified this property by repeating the experiment for all possible input modes in the chosen range. Barring experimental errors, the results are identical for different values of \(m\). The measured values of \(\langle s_{\Delta} \rangle\) for the \(m= 0\) case are presented in Fig. \ref{fig:comparison}(a) for the range of \(0 \leq \Delta  \leq 4\). The \(\Delta = 0\) line represents the fraction of the power that remains in the launched mode after propagation through atmospheric turbulence. The other curves indicate the portion of the power that has been transferred to the neighboring modes. As it can be seen, the \(\Delta = 0\) line begins at a value close to unity and drops steadily as the turbulence strength increases. The curves for all other values of \(\Delta\) initially increase as \(L/r_0\) increases and eventually decrease with a further increase of \(L/r_0\). The decrease at high turbulence levels occurs because the power in the input mode spreads among more and more PW modes. It can be seen that for a sufficiently large value of \(L/r_0\) the optical power spreads equally among all the modes.

\begin{figure}[h]
	\centerline{\includegraphics[scale=0.7]{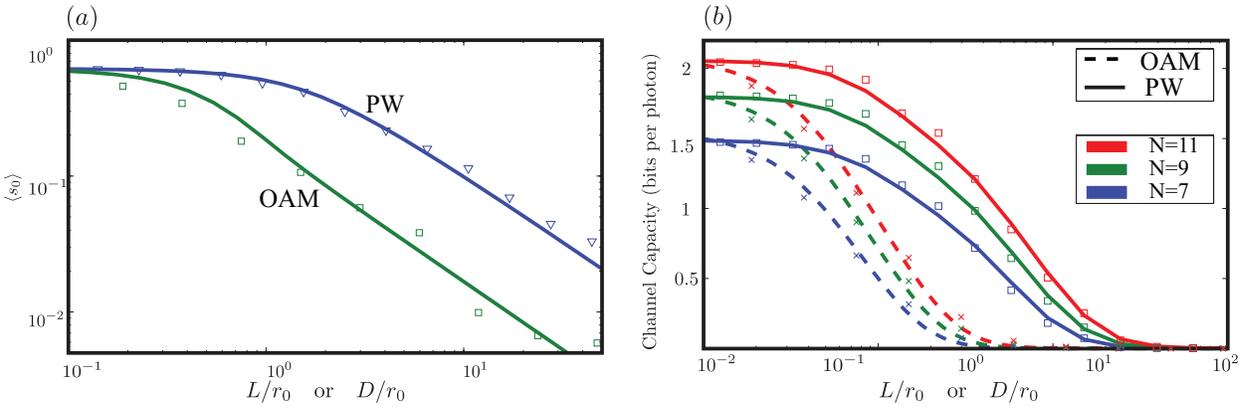}}
	\caption{(a) The fraction of energy \(s_0\) staying in an \(m = 0\) PW mode as it propagates through atmospheric turbulence is plotted along with the \(s_0\) values for an \(\ell = 0\) OAM mode. (b) Influence of Kolmogorov turbulence on the channel capacity of a PW communication channel as compared to that of an OAM channel. The channel capacities are calculated for a system dimensionality of \(N = 7, 9\) and \(11\). The solid lines present the theoretical predictions as fitted to the measured channel capacities at zero turbulence level.}
	\label{fig:DetailedComparison}
\end{figure}

We previously measured the performance of OAM modes for the same range of turbulence values \cite{rodenburg2012influence}. For comparison, the OAM results from this previous experiment are shown in Fig. \ref{fig:comparison}(b). It should be emphasized that the turbulence strength in this case is characterized by \(D/r_0\), where \(D\) is the diameter of the circular transmitting aperture. It is seen that the OAM cross-talk values behave in a qualitatively similar fashion to those of the PW modes. To make a quantitative comparison possible, the \(\Delta = 0\) values for OAM and PW modes are presented on the same plot in Fig. \ref{fig:DetailedComparison}(a). It can be seen that the power remains in the transmitted PW mode for larger levels of turbulence as compared to the OAM mode. More specifically, the PW curve reaches the same value as that of OAM at a turbulence level that is almost three times larger.

The consequences of these results for a communication link can be better understood using the concept of channel capacity. From the information theory, the channel capacity of a communication channel is defined as 
\begin{eqnarray}
C &  = & \max[H(x) - H(x\mid y)] \nonumber \\
     & = & \max_{\{Pi\}}\left[-\sum_i P_i \log_2 (P_i) + \sum_i P_i \sum_d P_{ij} \log_2(P_{ij})\right].	 
\end{eqnarray}
Here, \(N\) is the total number of modes in our system, \(P_i\) is the probability of transmission of mode \(i\), and \(P_{ij}\) is the conditional probability of transmission of mode \(i\) followed by detection of mode \(j\). We can calculate the channel capacity of our PW channel using the values of \(P_{ds} = \langle s_{(d - s)} \rangle\). Experimental data for the channel capacity of the PW communication channel with \(N= 7, 9\) and \(11\)  are shown in Fig. \ref{fig:DetailedComparison}(b). The channel capacity, C, is plotted as a function of the turbulence strength \(L/r_0\). For comparison, we have plotted the data for an OAM channel from Ref. \cite{malik2012influence} on the same figure. In both cases, the measured channel capacities are substantially lower than the theoretical limit of \(\log_2 N\) due to the limitations of the sorting techniques. It can be seen that the channel capacities for the PW modes tend to decrease much slower than those of the OAM modes as the turbulence level increases. More significantly, there exists a range of turbulence values for which the PW channel has non-zero capacities while the channel capacity of the OAM channel is almost equal to zero.

\section{Conclusions}
In summary, we have analyzed the performance of PW and OAM encodings in free-space communication systems employing spatial modes. Reviewing the theory of communication modes, we conclude that that the OAM basis set is the preferred encoding scheme when the effects of diffraction are dominant. In practical systems with large apertures where the effects of diffraction are negligible, the choice of encoding basis needs to be made considering the performance of the basis set under the effects of atmospheric turbulence. We have considered the PW basis as a candidate for such systems considering the simplicity of their generation and separation. The effects of Kolmogorov thin-phase turbulence on propagation of PW modes were studied experimentally. We have quantitatively compared the channel capacity of a PW-based communication system with the previous results from OAM-based systems.  Our results suggests the PW basis as the preferred encoding scheme for high Fresnel number communication systems.	 We would like to thank Stephen Barnett, David Miller, and Daniel Gauthier for helpful discussions. This work was supported by the DARPA/DSO InPho program and the Canadian Excellence Research Chair (CERC) program.
\bibliographystyle{tMOP}
%\bibliography{references}

\end{document}